\documentclass[12pt]{article}

\usepackage{paper2e}
\usepackage{mydefs2e}

\begin{document}

\newcommand{\ths}{\vartheta}

\newcommand{\N}{$\scr{N}=1$\xspace}

\begin{titlepage}
\preprint{UMD-PP-02-019 \\
JHU-TIPAC-2001-05}

\title{Anomaly Mediated Supersymmetry Breaking
      \\\medskip
in Four Dimensions, Naturally}


\author{Markus A. Luty}

\address{Department of Physics, University of Maryland\\
College Park, Maryland 20742, USA\\
{\tt mluty@physics.umd.edu}}

\author{Raman Sundrum}

\address{Department of Physics and Astronomy, Johns Hopkins University\\
Baltimore, Maryland 21218, USA\\
{\tt sundrum@pha.jhu.edu}}

\begin{abstract}
We present a simple four-dimensional
model in which anomaly mediated supersymmetry breaking naturally dominates.
The central ingredient is that the hidden sector is near a strongly-coupled
infrared fixed-point for several decades of energy below the Planck scale.
Strong renormalization effects then sequester the hidden sector from the
visible sector.
Supersymmetry is broken dynamically and requires no small input parameters.
The model provides a natural and economical explanation of the hierarchy
between the supersymmetry-breaking scale and the Planck scale, while
allowing anomaly mediation to address the phenomenological challenges posed
by weak scale supersymmetry.
In particular, flavor-changing neutral currents are naturally near their
experimental limits.
\end{abstract}

\end{titlepage}

\section{Introduction}
Anomaly mediated supersymmetry breaking (AMSB) \cite{amsb1, amsb2} is a
general supergravity mechanism that is tightly constrained by local
supersymmetry.
AMSB may play an important role in solving the major phenomenological
problems of weak scale supersymmetry (SUSY):
the flavor, $\mu$, and gaugino mass problems.
For example, \Ref{pomrat} describes a complete and very plausible extended
supersymmetric standard model where anomaly mediation is the main
ingredient in solving these problems and leads to a realistic and
distinctive spectrum.
\Refs{amsbfix} describe other proposals for weak scale AMSB.

In order for AMSB to dominate in the observable sector, SUSY breaking must
originate in a special type of hidden sector.
A general hidden  sector model has the form
\beq[genhid]
{\cal L} = {\cal L}_{\rm SUGRA} + {\cal L}_{\rm visible}
+ {\cal L}_{\rm hidden} +
       {\cal L}_{\rm mixed},
\eeq
where the first three terms are self-explanatory, while ${\cal L}_{\rm mixed}$
contains Planck-suppressed terms involving both visible and hidden
fields that cannot naturally be forbidden by symmetries.
AMSB in the visible sector arises from minimal coupling to supergravity,
in particular the auxiliary scalar field in the minimal formulation.
By supercovariance, this scalar couples via visible mass scales, in particular
the renormalization scale associated with the scale anomaly in radiative
corrections (hence `anomaly mediation').
Therefore, supersymmetry breaking effects arising from AMSB are suppressed
by loop factors.
In general hidden sector models, larger visible SUSY breaking can arise
directly from the hidden  sector through terms in ${\cal L}_{\rm mixed}$.
Therefore, in order for AMSB to dominate ${\cal L}_{\rm mixed}$ must be
strongly suppressed.%
\footnote{Note that this is true in superspace, but not after component
level field redefinitions to go to Einstein frame.
See \Ref{pomrat}.}
That is, the hidden and visible sectors are `sequestered'.

In \Refs{amsb1, radstab} it was shown that sequestering can be achieved if
the visible and hidden sectors are localized on different 3-branes
separated in extra dimensions.
Recently, we demonstrated that highly warped supersymmetric anti de Sitter
space (AdS) compactifications could be stabilized with sufficient
sequestering \cite{susywarp}.
AdS/conformal field theory (CFT) duality \cite{adscft} applied to such
compactifications \cite{adscftrs} then suggests that sequestering can also
arise in a purely 4D context with the help of strongly coupled conformal
dynamics.
In \Ref{last}, we showed that sequestering in fact occurs in a large
class of supersymmetric CFT's.
  We also presented a specific model incorporating SUSY breaking
of the required type.
This model is technically natural, but it requires several unexplained
small numbers.
In this paper, we will present a very simple and plausible model of
conformal sequestering, in which all large hierarchies are dynamically
generated.
Using `\naive dimensional analysis' to estimate the strong interaction
coefficients, we find that the model easily gives enough sequestering
so that anomaly mediation dominates, and 
flavor-changing neutral currents are near their experimental limits.

The basic structure of our model is as follows.
The central component of ${\cal L}_{\rm hid}$ is a SUSY theory that is near
a strongly-coupled conformal fixed point below the Planck scale.
The infrared approach to the fixed point is governed by an order one
critical exponent $\beta_*'$.
Imposing certain exact hidden symmetries restricts the hidden sector
factors in ${\cal L}_{\rm mixed}$ to have the same form as the operators
in ${\cal L}_{\rm hid}$.
Because of this, the operators in ${\cal L}_{\rm mixed}$ can be viewed
as perturbations of hidden sector couplings with visible sector coefficients.
All such perturbations are suppressed by $(\mu/M)^{\beta_*'}$ as the hidden
sector approaches the fixed point, where  $\mu$ is the renormalization
scale and $M$ is the Planck scale.
This is the conformal sequestering mechanism.

Superconformal field theories naturally have a moduli space.
They are exactly superconformal only at the origin of moduli space, but
away from the origin, superconformal invariance is spontaneously broken.
The degeneracy of these vacua is lifted by weak, even technically
irrelevant perturbations to the fixed point theory.
We use such effects to generate an effective potential for the hidden
moduli space which stabilizes the moduli away from the origin with a SUSY
breaking vacuum energy.
The small numbers needed to ensure that the SUSY breaking scale and the
moduli VEVs are hierarchically smaller than the Planck scale are naturally
generated by non-perturbative 
effects.

For an earlier application of strong conformal dynamics
(in the visible sector) to supersymmetric model-building see
\Ref{nelson-strassler}.

\section{The Model}

In this paper, we focus on  the hidden sector
   and the mechanism for sequestering from the visible sector.
The visible sector can be any theory for which AMSB yields an acceptable
phenomenology.
For now, we will restrict ourselves to global SUSY. In Section 4,
we will
%
%
consider the
SUGRA corrections to the effective potential,
%
%
which are important for modulus stabilization and cancelling the
cosmological constant.

Our model of the hidden sector consists of two supersymmetric QCD (SQCD)
   subsectors:
a $SU(2)$ gauge theory with 4 flavors (8 fundamentals) $T^{J a}$
($J = 1, \ldots, 4$; $a = 1, 2$), denoted by SQCD$_2$;
and a $SU(3)$ gauge theory with 2 flavors $P^a$, $\bar{P}_a$
($a = 1, 2$), denoted by SQCD$_3$.
%
Throughout the paper we will suppress all gauge indices, and we will
suppress the $a = 1, 2$ index when the meaning is clear.
%
We impose the following symmetries on the hidden sector:
permutations of the $T^{J}$, multiplication
of any of the $T^J$ by $-1$, charge conjugation for SQCD$_3$, and a
global $SU(2)$ symmetry acting on the $a = 1, 2$ index.%
\footnote{For readers concerned by quantum gravity violation of global
symmetries: the $SU(2)$ group can be weakly gauged, or can be replaced by a
suitable discrete subgroup.}
The theory has a superpotential invariant under these symmetries:
\beq[Wtree]
W = \frac{\la}{M} \sum_J (T^J T^J) (\bar{P} P)
+ \frac{\la'}{M} \sum_{J \neq K} (T^J T^J) (T^K T^K),
\eeq
where $M$ is the Planck scale.
We will show that this simple model sequesters itself from the visible
sector and has a local minimum that dynamically breaks SUSY.

The SQCD$_2$ sector is at the self-dual point of Seiberg's conformal window
\cite{Seiberg} and we will assume that it starts near its IR fixed point
coupling at the Planck scale.
It is therefore strongly coupled.
We assume the SQCD$_3$ sector is weakly coupled at the Planck scale.
We also assume that the superpotential couplings $\la$ and $\la'$ are
sufficiently small that they can be treated as perturbations of the
SQCD$_2$ fixed point.

The leading dangerous terms in ${\cal L}_{\rm mixed}$ compatible
with the hidden sector symmetries are
\beq[Lmixedops]
{\cal L}_{\rm mixed}(M) = \myint d^4 \th \left[
\frac{c^j{}_k}{M^2} Q^{\dagger}_j Q^k
\sum_{J} T_J^\dagger T^J + \frac{(c_P)^j{}_k}{M^2} Q^{\dagger}_j Q^k
(P^\dagger P + \bar{P}^\dagger \bar{P}) \right],
\eeq
where the $Q_j$ are visible chiral superfields.
The danger is that for $c$, $c_P$ of order unity and containing
SM flavor violation, flavor violating visible scalar masses will be generated
upon SUSY breaking in the hidden sector that will dominate over the
flavor-blind AMSB contributions.
We will show that the SQCD$_2$ conformal dynamics naturally suppresses the
effects of $c$, $c_P$ at low energies, allowing AMSB
to dominate the visible sector.
Supergravity loops can contribute to
mixed couplings, but they are dominant in the ultraviolet, so their
leading effects can be absorbed into the $c$ coefficients.


\newpage
\section{Sequestering}


We first consider the limit $\la = \la' = 0$.
In this limit the SQCD$_3$ sector completely decouples and we can
omit it from the discussion.
The leading terms in $\scr{L}_{\rm mixed}$ (see \Eq{Lmixedops})
can be viewed as perturbations to the wavefunction of the
hidden fields renormalized at the Planck scale\footnote{
Note that without imposing the hidden-flavor symmetries discussed in
Section 2,
the mixed terms could be more general than this form. In this case
we would encounter the difficulties discussed in \Ref{last}.}:
\beq[hiddtheory]
(\scr{L}_{\rm hidden} + \scr{L}_{\rm mixed})(M)
= \myint d^4\th\, Z_0 T^\dagger T +
\left( \myint d^2\th\, \tau_{\rm hol,0} \tr W^\al W_\al + \hc \right),
\eeq
where
\beq[UVpert]
Z_0 = z_0 + \frac{c^j{}_k}{M^2} Q^\dagger_j Q^k.
\eeq
We will explain the role of 
$z_0$ below.
$\tau_{\rm hol,0}$ is the holomorphic $SU(2)$ gauge coupling.
The theory defined by
\Eq{hiddtheory} has only one physical coupling, namely the physical
gauge coupling $\tau = 1/g^2$, given by%
\footnote{Note that $Z_0$ in \Eq{UVpert} is a vector superfield.
By `analytic continuation into superspace' \Eq{NSVZ} can be interpreted
as an equality of vector superfields \cite{analcont}.}
\beq[NSVZ]
\tau = \Re(\tau_{\rm hol}) - \frac{F}{8\pi^2} \ln Z + \frac{N}{8\pi^2} \ln\tau
+ f(\tau).
\eeq
where $N$ is the number of colors and $F$ is the number of flavors;
in our theory, $N = 2$, $F = 4$.
Here $f(\tau) =  \hbox{\rm constant} + \scr{O}(\tau^{-1})$
parameterizes the scheme dependence.
The perturbation \Eq{UVpert} (holding $\tau_{\rm hol, 0}$ fixed)
therefore gives rise to a perturbation of the physical gauge coupling.
Taking the derivative $d/dt$  of \Eq{NSVZ}, where $t \equiv \ln\mu / M$,
we obtain
\beq[NSVZbeta]
\be(\tau) = \frac{\displaystyle \frac{b}{8\pi^2}
- \frac{F}{8\pi^2} \ga(\tau)}
{\displaystyle
1 - \frac{N}{8\pi^2} \frac{1}{\tau}
- f'(\tau)},
\qquad b = 3N - F,
\eeq
where $\be \equiv d\tau/dt$, $\ga \equiv d\ln Z / dt$, and we have used
$d\tau_{\rm hol} / dt = b / 8\pi^2$.
In the `NSVZ scheme' $f \equiv 0$, \Eqs{NSVZ} and \eq{NSVZbeta} are
the famous formulae of \Refs{NSVZ}.

Because \Eq{UVpert} is a perturbation to the UV gauge coupling,
it is clear that it is irrelevant near the IR fixed point.
This means that the effects of the perturbation
$c^j{}_k Q^\dagger_j Q^k / M^2$
are suppressed in the IR.
This is the underlying mechanism for sequestering in this class of models
\cite{last}.

We now make this quantitative.
Exactly at the fixed point, $\tau = \tau_* = $~constant,
so $\ga(\tau) = \ga(\tau_*) \equiv \ga_* = $~constant.
Therefore (taking $Z_*(t = 0) = 1$)
\beq[Z*]
Z_*(t) = e^{\ga_* t}.
\eeq
The theory is at a fixed point despite the running of $Z$ because the
running of $\tau_{\rm hol}$ compensates so that
$\be(\tau_*) = 0$.
     From \Eq{NSVZbeta} we see that this requires \cite{Seiberg}
\beq
\ga_* = \frac{b}{F}.
\eeq

We now consider the perturbations about the fixed point.
We expand the RG functions in $\De\tau \equiv \tau - \tau_*$
to first order
to define critical exponents
\beq[fixedexp]
\be(\tau) &\simeq \be_*' \cdot \De\tau, 
\\
\ga(\tau) &\simeq \ga_* + \ga'_* \cdot \De\tau. 
\eeq
We factor out the fixed point running by defining
\beq
\De\ln Z \equiv \ln Z - \ga_* t.
\eeq
Then we have
\beq
\frac{d (\De\ln Z)}{dt} = \ga'_* \cdot \De\tau,
\qquad
\frac{d (\De\tau)}{dt} = \be'_* \cdot \De\tau.
\eeq
Because of the relation \Eq{NSVZ}, these equations are not independent.
Using \Eq{NSVZ} we can write an RG equation for $\De \ln Z$ alone:
%
%
\beq[ZtildeRG]
\frac{d (\De\ln Z)}{dt} = \be'_* \left[ \De\ln Z
- \frac{8\pi^2}{F} \De\tau_{\rm hol,0} \right].
\eeq
Here $\De\tau_{\rm hol} \equiv \tau_{\rm hol} - \tau_{\rm hol,*}$, and
$\tau_{\rm hol,*}$ satisfies \Eq{NSVZ} for $\tau = \tau_*, Z = Z_*$.
The deviation from the fixed point in the UV is parameterized by
$Z_0 \ne 1$ (see \Eq{UVpert} and \Eq{Z*}) and $\De\tau_{\rm hol,0} \ne 0$.
     From \Eq{NSVZ} we can see that these are not independent perturbations,
so we can choose $\De\tau_{\rm hol,0} = 0$ and parameterize the
perturbation by $Z_0$ alone.
The solution to \Eq{ZtildeRG} is then simply
\beq[delZ]
      \De\ln Z
= e^{\be'_* t}
      (\De\ln Z)_0.
\eeq
$\beta_*'$ is a strong-interaction critical exponent of order one.
Since the dangerous terms in $\scr{L}_{\rm mixed}$
are contained in $\De\ln Z_0$,
this clearly shows the sequestering.

%

Now we include the effects of $\la$, $\la'$, and $\tau_3$.
     From now on we specialize to the case $F = 2N$ for the
SQCD$_2$ sector, so that $\ga_* = \frac 12$.
We must now include the additional mixed terms
\beq[ZP]
Z_{P,0} = 1 + \frac{(c_P)^j{}_k}{M^2} Q^\dagger_j Q^k.
\eeq
Because the $SU(3)$ sector is not a CFT, we expect at most an order 1
renormalization of $Z_P$.
Since we are only interested in the order of magnitude of $Z_P$,
we will simply use the approximation $Z_P \simeq Z_{P,0}$.
The mixed terms in \Eq{ZP} do not directly give rise to large
visible soft masses because in our model the dominant source of
SUSY breaking is in the SQCD$_2$ sector.

However, we must determine the leading effects of the perturbation
\Eq{ZP} on the SQCD$_2$ sector.
These can be studied in the RG equation for $\De\ln Z$:
\beq[ZtildeRGlambdageneral]
\frac{d (\De\ln Z)}{dt}
= \ga'_* \cdot \De\tau
+ \De\ga(\tau, \tau_3, \la_{\rm phys}, \la'_{\rm phys}),
\eeq
where
\beq[lambdas]\bal
\la_{\rm phys} &= \frac{\la \mu}{M Z Z_P}
= \frac{\la e^{t/2} e^{-\De\ln Z}}{Z_P},
\\
\la'_{\rm phys} &= \frac{\la' \mu}{M Z^2}
= \la' e^{-2\De\ln Z}.
\eal
\eeq
While $\De\ga$ is a small perturbation in \Eq{ZtildeRGlambdageneral},
it becomes comparable to the first term on the \rhs in the IR.
We must show that this does not spoil sequestering.
Since we are expanding around the fixed point we can set
$\tau = \tau_*$ in $\De\ga$.
We will use \Eq{ZtildeRGlambdageneral} only in the regime where
the SQCD$_3$ sector is unbroken and weakly coupled.
In this regime, we can neglect the running due to $\tau_3$.
The leading terms are therefore
\beq[ZtildeRGlambda]
\frac{d (\De\ln Z)}{dt}
= \ga'_* \cdot \De\tau
+ \frac{ |\la_{\rm phys}|^2}{\rho^4}
+ \frac{ |\la'_{\rm phys}|^2}{\rho^4}.
\eeq
Because of the SQCD$_2$ strong interaction uncertainties, we cannot
compute the coefficients of the last two terms precisely, but we have
estimated their order of magnitude using  `naive dimensional analysis' (NDA)
\cite{nda,SUSYNDA}. Here, and later in the paper, we will
give our NDA estimates in terms of
\begin{equation}
\rho \sim 4 \pi.
\end{equation}
Separate order one uncertainties should then be ascribed to
   different terms,
but these will not be written explicitly.

Once again, we would like to use \Eq{NSVZ} to eliminate $\De\tau$ on the
\rhs of \Eq{ZtildeRGlambda} in favor of $\De\ln Z$.
In the presence of the additional couplings $\la$, $\la'$, and $\tau_3$
\Eq{NSVZ} remains true, but the scheme dependent function $f$ is in
general a function of all the couplings.
However, we can always choose a scheme where $f$ is a function of $\tau$
alone.
In such a scheme we have
\beq[schemeZtildeeq]
\frac{d (\De\ln Z)}{dt}
= \be'_* \De\ln Z
+ \frac{|\la_{\rm phys}|^2}{\rho^4}
+ \frac{ |\la'_{\rm phys}|^2}{\rho^4}.
\eeq
The last two terms on the right-hand side are subdominant perturbations
compared to the first term unless $\De\ln Z$ is small. Therefore we can
approximate the last two terms using \Eq{lambdas} in the limit
$\De\ln Z \rightarrow 0$.
Also $Z_P$ runs only perturbatively, so we can approximate
$Z_P \simeq Z_{P,0}$.
We then obtain the approximate solution
\beq[DelogZ]
\bal
\De \ln Z &\simeq e^{\be'_* t} (\De\ln Z)_0
+ \frac{ |\la|^2}{\rho^4 Z_{P,0}^2}
\, \frac{e^{\be'_* t} - e^t}{\be'_* - 1}
\\
&\qquad
+ \frac{ |\la'|^2}{\rho^4} \, \frac{e^{\be'_* t} - 1}{\be'_*}.
\eal\eeq
The first two terms contain mixed terms, but are sequestered,
while the third term is not sequestered, but contains no mixed
terms.
Therefore, all mixed terms are suppressed in this model provided
that there is a sufficiently large range of scales for which the
SQCD$_2$ sector is near the fixed point.
In fact, the above perturbations due to $\lambda, \lambda'$ have
subdominant effects to   others we will later identify and to $\De\ln Z$.

It is convenient to summarize the RG near the fixed point
by writing the effective lagrangian
\beq\bal
\scr{L} &\simeq \myint d^4\th \left[
\mu^{1/2}
\! (1 + \De \ln Z) \, \tilde{T}^\dagger \tilde{T}
+ Z_{P,0} \, (P^\dagger P + \bar{P}^\dagger \bar{P}) \right]
\\
& \quad + \myint d^2\th \left[
      \frac{\la}{M^{1/2}} \sum_J (\tilde{T}^J \tilde{T}^J) (\bar{P} P)
+ \la' \sum_{J \neq K} (\tilde{T}^J \tilde{T}^J)
(\tilde{T}^K \tilde{T}^K)
\right] + \hc
\\
&\quad + \hbox{\rm gauge\ kinetic\ terms},
\eal\eeq
where we have defined the rescaled fields \cite{last},
\beq
\tilde{T} \equiv \frac{T}{M^{1/4}}.
\eeq
This rescaling removes the leading $M$ dependence of the lagrangian, and
makes the canonical dimension of the $\tilde{T}$ fields the same as their
fixed-point scaling dimension in chiral operators.

\section{Supersymmetry Breaking}
We now determine the vacuum in this theory.
We will show that there is a locally stable vacuum with broken SUSY at
$T \ne 0$.

In the absence of the superpotential couplings \Eq{Wtree}, the
SQCD$_2$ theory has 13 independent moduli, which can be parameterized
by the $SU(2)$ gauge invariant `meson' operators of the form $T^{Ja}
T^{Kb}$ subject
to classical constraints.
Away from the origin of moduli space the superpotential couplings
proportional to $\la'$ reduce the moduli space to a single flat direction,
which we assume is in the direction%
\footnote{For more detail on the moduli space of this theory, see \Ref{last}.}
\beq
TT \propto \pmatrix{ X^{3/4} \, \ep & 0 \cr 0 & 0 \cr},
\qquad
\ep = \pmatrix{0 & 1 \cr -1 & 0 \cr},
\eeq
where we use the basis
\beq
T = \pmatrix{ T^{11} \cr T^{12} \cr \vdots \cr T^{42} }.
\eeq
The field $X$ parameterizes the flat direction.
A VEV for $X$ breaks the conformal symmetry, so $X$ is the Nambu-Goldstone
mode for spontaneous breaking of scale symmetry.
We have defined $X$ so that it has dimensions of mass.

The first threshold in this theory is given by the VEV $\avg{X}$, where the
conformal symmetry is spontaneously broken.
NDA tells us that the physical threshold is at a scale
$\sim
 (\rho \avg{\tilde{T}})^{4/3}$, and that the canonically normalized modulus
field
is $X \sim \rho^{1/3} \tilde{T}^{4/3}$.
The effective lagrangian below the scale of conformal symmetry breaking is
written in terms of the modulus $X$ and the SQCD$_3$ fields:
\beq[LeffX]
\bal
\!\!\!\!\!\!\!
\scr{L}_{\rm eff}(\mu \lsim \rho |X|)
&\sim \myint d^4\th \left\{ \left[
1 + \De\ln Z(\mu \sim \rho |X|) \right] X^\dagger X
+ Z_{P,0} \, (P^\dagger P + \bar{P}^\dagger \bar{P}) \right\}
\\
&\qquad
+ \myint d^2\th\, \frac{\la}{\rho^{1/2} M^{1/2}} X^{3/2} \bar{P} P + \hc
\\
&\qquad
+ SU(3)\ \hbox{\rm gauge\ kinetic\ terms}.
\eal\eeq

The superpotential in \Eq{LeffX} gives rise to a mass for the $P$ fields
\beq
m_P \sim \frac{\la}{\rho^{1/2} M^{1/2}} \avg{X}^{3/2}.
\eeq
We consider the case $m_P > \La_3$, where $\La_3$ is the scale
where the SQCD$_3$ gauge theory with 2 flavors becomes strong, and
will check the self-consistency of this choice later.
In this case, we can integrate out the $P$ fields perturbatively at
the scale $m_P$, and the effective theory is
\beq[LeffmP]
\bal
\scr{L}(\mu \lsim m_P) &\sim \myint d^4\th \left[
1 + \De\ln Z(\mu \sim \rho |X|)
\vphantom{\left( \frac{\rho |X|}{m_P} \right) }
\right. \\
& \qquad\qquad\quad
+ \left. \frac{\la^2}{\rho^{3} \, Z_{P,0}^2 M} \,
|X| \ln \left( \frac{\rho |X|}{m_P} \right)
\right] X^\dagger X
\\
& \qquad
+ SU(3)\ \hbox{\rm gauge\ kinetic\ terms}.
\eal\eeq
The ln$|X|$ term gives the leading effect of $P$ loops between the scales
$\rho \avg{X}$ and $m_P$.\footnote{
The precise coefficient of this log term is calculable but
is unimportant because of the order 1
uncertainties in the other coefficients such as $m_P$.}

The pure $SU(3)$ gauge theory becomes strong at a scale%
%
\beq
\La_{\rm 3,eff} \sim m_P^{2/9} \La_3^{7/9}.
\eeq
Gaugino condensation gives rise to an effective superpotential
\beq[Polonyi]
W_{\rm dyn} \sim \frac{\Lambda_{\rm 3,eff}^3}{\rho^2} \sim \La_{\rm int}^2 X,
\eeq
where
\beq[intscale]
\La_{\rm int}^2 \sim \frac{\la^{2/3} \La_3^{7/3}}{\rho^{7/3} M^{1/3}}.
\eeq
\Eq{Polonyi} is the superpotential of a Polonyi model,
which breaks SUSY provided that the \Kahler terms stabilize the
field $X$.
The vacuum energy is then of order $\La_{\rm int}^4$, and therefore
$m_{3/2} \sim \La_{\rm int}^2 / M$.

The effective potential for $X$ including the \Kahler terms of
\Eq{LeffmP} is
\beq
\nonumber
\!\!\!\!\!\!\!
V_{\rm eff} &=
\La_{\rm int}^4 \left/ \frac{\partial^2 K_{\rm eff}}{\partial
X\partial X^\dagger}\
\right.
+ \De V_{\rm SUGRA}
\\
&\sim \La_{\rm int}^4 \left[
1 + (\ln z_0) \left( \frac{\rho |X|}{M} \right)^{\be'_*}
- \frac{\la^2}{\rho^{3}} \, \frac{|X|}{M} \,
\ln \left( \frac{\rho |X|}{m_P} \right) \right]
\nonumber\\
&\qquad\quad
   - \Lambda_{\rm int}^4 (1 + \mathop{\rm Re}(X)/M).
\eeq
Here we have written out the leading terms in the solution for
$\De\ln Z$ (see \Eq{DelogZ}) and used $(\De \ln Z)_0 = \ln z_0$
(see \Eq{UVpert}). We have dropped terms comparable to the $\ln|X|$ term
that are not log-enhanced.
The coefficient of the $\ln z_0$ term depends on strong interactions,
but we can choose the sign of $\ln z_0$ so that the sign of this term
is positive. The last line contains the leading SUGRA corrections
once we add a constant superpotential so as to
cancel the $\Lambda_{\rm int}^4$ contribution to the cosmological
constant.
We will demand
   that the supergravity corrections to the potential dominate over
   the $|X| \ln|X|$ term. This gives the restriction
\beq[SUGRAconst]
\frac{\la^2}{\rho^{3}} \ln\left(
\frac{\rho \avg{X}}{m_P} \right) \lsim 1.
\eeq
%
We then find a stable minimum at
\beq[paramseq]
\hbox{\rm sequestering}
\equiv \left( \frac{\rho \avg{X}}{M} \right)^{\be'_*}
\sim \left[\frac{1}{\rho \ln z_0}
\right]^{\be'_* / (\be'_* - 1)},
\eeq
where we have solved for the sequestering factor for the mixed terms in
$Z_0$.
(By \Eq{SUGRAconst}, the dangerous mixed terms arising from
$Z_{P, 0 }$ are even more suppressed.)
The term $\ln z_0$ parametrizes the deviation of SQCD$_2$ from the
fixed point at the Planck scale, and must be small enough that we can trust
the fixed-point expansions, \Eq{fixedexp}.
NDA yields tells us that this requires $\ln z_0 \lsim 1$.
The anomalous dimension $\be'_*$ is order 1 (and positive), and therefore
the sequestering factor is an order-1 power of a loop suppression factor
(up to a logarithmic correction).

In fact, there is an adjustable parameter that controls the amount of
sequestering in our model.
It is completely natural for the SQCD$_2$ sector to enter the
strong-coupling conformal regime at a sub-Planckian scale, $\tilde{M} < M$,
although we have taken the two scales to be equal.
In this more general case, we must substitute $1/\rho \rightarrow
\tilde{M}/(\rho M)$ on the right-hand side of \Eq{paramseq}.
We can therefore obtain any desired amount of sequestering by taking
$\tilde{M} \ll M$.
Our analysis assumed that $\Lambda_3 < m_P$, so that the $P$'s were integrated
out of the theory before the SQCD$_3$ subsector became strongly coupled.
This naturally occurs for sufficiently small $\Lambda_3$, which also sets
the SUSY breaking scale according to \Eq{intscale}.
At the qualitative level, these observations show that the model naturally
breaks SUSY far below the Planck scale and generates a large amount of
sequestering.
In the next section we will see that quantitatively, we must saturate the
inequalities \Eq{SUGRAconst}, $\Lambda_3 < m_P$, and $\ln z_0 \lsim 1$ in
order to get maximal sequestering for the real world.
It is also optimal to take $\tilde{M} \sim M$ as we have throughout the
paper.



\section{Numerical Estimates}

We now turn to the numerical estimates in this model.
Using \Eq{intscale} and  \Eq{paramseq} (with $\ln z_0 \sim 1$)
to eliminate the dependence
on $\Lambda_3$ and $\avg{X}$, the constraint $\Lambda_3 < m_P$ can be
written
\beq[bd]
\bigl( \hbox{\rm sequestering} \bigr)^{7/2} &\gsim
\frac{\rho^{7/2}}{\lambda^3} \left( \frac{\La_{\rm int}}{M} \right)^2.
\eeq
%
%
We see that we obtain maximal sequestering by saturating the bound
\Eq{SUGRAconst}.
We will approximate the logarithm in \Eq{SUGRAconst} as order one.
Note that the resulting $\lambda \sim \rho^{3/2}$ is smaller than the
strong-coupling value, $\lambda_{\rm strong} \sim \rho^2$.
Substituting into \Eq{bd} then gives a bound on the sequestering factor:
\beq
{\rm sequestering}  \gsim
\frac{1}{\rho^{2/7}}
\left( \frac{\La_{\rm int}}{M} \right)^{4/7}
\sim  6 \times 10^{-5}.
\eeq
We have taken $M = 2.4 \times 10^{18} \GeV$ and
$\La_{\rm int} \simeq 3 \times 10^{11}\GeV$.
By \Eq{paramseq}, this maximal level of sequestering is obtained for
$\be'_* \simeq 1.2$.
The minimum is at $\avg{X} \sim 10^{14}\GeV$,
and the mass of $X$ is of order $5 \times 10^6\GeV$.

The amount of sequestering
is sufficient for AMSB to dominate in the visible sector,
and is within an order of magnitude
of the sequestering factor $3 \times 10^{-6}$ \cite{last} \cite{fcnc}
required to adequately suppress CP-conserving flavor violation
in anomaly-mediated SUSY breaking if the coefficients $c$ of
\Eq{Lmixedops} are
of order $1$. Given the considerable
uncertainties in the  strong-interaction coefficients,
our maximal sequestering could easily be at or below this
flavor-violation bound. Of course it is also possible that the $c$'s of
\Eq{Lmixedops} are of order $1/10$.



We now consider briefly the cosmology of this model.
In general, models of the hidden sector suffer from the Polonyi
problem \cite{Polonyi}.
Briefly stated, the problem is that models with moduli generally
have a cosmological epoch where coherent oscillations of the moduli
dominate the energy density of the universe, and the interactions
of the moduli with the visible sector are too weak to reheat the
universe to a sufficiently high temperature to allow nucleosynthesis.
In the present model, this problem is less severe than in
standard hidden sector models because the mass of the modulus is large
compared to the weak scale and the self-interactions of the moduli are
much
stronger than gravitational strength. We will leave a full analysis of this
issue for future work.


Another cosmological issue is the fact that the minimum we have
found is a false vacuum.
There is a supersymmetric vacuum at the origin $T = 0$, but
because $\avg{X} \gg \La_{\rm int}$, the tunnelling rate is
suppressed by a large exponent and is cosmologically safe \cite{cosmo}.


\section{Discussion and Conclusions}
It is remarkable that the simple four-dimensional model of the hidden
sector presented here dynamically breaks supersymmetry and sequesters
itself from the visible sector, naturally allowing anomaly mediation to
dominate visible sector supersymmetry breaking.
We  believe that similar mechanisms of sequestering and dynamical
supersymmetry breaking can occur in a large class of  models,
although it is difficult to check this outside of supersymmetric QCD
because of the limited number of
superconformal theories that are known explicitly.

According to our estimates,
CP conserving flavor-changing neutral current processes are near their
experimental limits.
Given the large uncertainties, it is possible that there is more
sequestering than given in our estimates, so that CP-violating flavor
violation is also sufficiently suppressed.
Alternatively, suppressing CP violating flavor violation may require
additional structure.
In any case, we expect some flavor-changing neutral current processes
to be close to their experimental limits.


We hope that this work will help open new directions
for constructing complete, compelling, and realistic hidden sector
models of supersymmetry breaking.

\section*{Acknowledgements}
We thank the Fermilab theory group for hospitality during
the initial stage of this work.
M.A.L. was supported by NSF grant PHY-98-02551. R.S. was supported in
part by a DOE Outstanding Junior Investigator award DEFG0201ER41198
and in part by NSF Grant PHY-0099468.

\newpage

\end{document}